\documentclass[twocolumn,preprintnumbers,nofootinbib,prd]{revtex4}
\usepackage{psfrag,graphicx,epsfig,amsmath,amssymb,bm}
\begin{document}

\preprint{FERMILAB-PUB-09-002-T, MZ-TH/09-01} 

\title{\boldmath
Infrared singularities of scattering amplitudes in perturbative QCD  
\unboldmath}

\author{Thomas Becher\,$^a$ and Matthias Neubert\,$^b$} 

\affiliation{$^a$\,Fermi National Accelerator Laboratory,
P.O. Box 500, Batavia, IL 60510, U.S.A.\\
$^b$\,Institut f\"ur Physik (THEP), Johannes Gutenberg-Universit\"at,
D-55099 Mainz, Germany}
\date{\today}

\begin{abstract}
\noindent
An exact formula is derived for the infrared singularities of dimensionally regularized scattering amplitudes in massless QCD with an arbitrary number of legs, valid at any number of loops. It is based on the conjecture that the anomalous-dimension matrix of $n$-jet operators in soft-collinear effective theory contains only a single non-trivial color structure, whose coefficient is the cusp anomalous dimension of Wilson loops with light-like segments. Its color-diagonal part is characterized by two anomalous dimensions, which are extracted to three-loop order from known perturbative results for the quark and gluon form factors. This allows us to predict the three-loop coefficients of all $1/\epsilon^k$ poles for arbitrary $n$-parton scattering amplitudes, generalizing existing two-loop results.
\end{abstract}

\maketitle

Perturbative expressions for quark-gluon scattering amplitudes in massless QCD contain infrared (IR) singularities, which originate from configurations where loop momenta become soft or collinear. In predictions for physical observables these cancel against corresponding singularities from real gluon emission by virtue of the KLN theorem \cite{Kinoshita:1962ur}. Nevertheless, the IR singularities are interesting in their own right. After the cancellation of IR singularities logarithmic terms remain, which depend on the phase-space cuts imposed. For high-energy scattering processes with low-mass jets in the final states these Sudakov logarithms dominate the cross section. An understanding of the structure of IR singularities can be used to predict and resum these logarithmically enhanced contributions to all orders. It also serves as a consistency check on loop calculations.

The IR singularities of QED are understood to all orders in perturbation theory. They arise from multiple soft photon emissions. Eikonal identities ensure that higher-order soft radiation is obtained simply by exponentiating the leading-order contribution \cite{Yennie:1961ad}. For non-abelian gauge theories the situation is more complicated. The soft emissions receive genuine higher-order corrections \cite{Frenkel:1984pz}, and in addition to soft singularities also collinear ones arise. While factorization proofs guarantee their absence in inclusive observables \cite{Collins:1989gx}, an all-order result for the IR singularities of QCD amplitudes is currently still lacking. An important step toward this goal was made by Catani \cite{Catani:1998bh}, who correctly predicted the singularities of two-loop amplitudes apart from the $1/\epsilon$ pole term. His formula states that the product 
\begin{equation}
   \left[ 1 - \frac{\alpha_s}{2\pi}\,\bm{I}^{(1)}(\epsilon)
    - \left( \frac{\alpha_s}{2\pi} \right)^2 \bm{I}^{(2)}(\epsilon)
    \right] |{\cal M}_n(\epsilon,\{p\})\rangle \,,
\end{equation}
where $\alpha_s\equiv\alpha_s(\mu)$, and $|{\cal M}_n(\epsilon,\{p\})\rangle$ with $\{p\}\equiv\{p_1,\dots,p_n\}$ denotes an ultraviolet (UV) renormalized, on-shell $n$-parton scattering amplitude with IR singularities regularized in $d=4-2\epsilon$ dimensions, is free of IR poles through ${\cal O}(\alpha_s^2)$. The subtraction operators $\bm{I}^{(n)}(\epsilon)\equiv \bm{I}^{(n)}(\epsilon,\{p\},\mu)$ are defined as
\begin{eqnarray}\label{I2}
   \bm{I}^{(1)}(\epsilon) 
   &=& \frac{e^{\epsilon\gamma_E}}{\Gamma(1-\epsilon)}
   \sum_i \left( \frac{1}{\epsilon^2} + \frac{g_i}{\bm{T}_i^2}\,
   \frac{1}{\epsilon} \right) \!
   \sum_{j\neq i} \frac{\bm{T}_i\cdot\bm{T}_j }{2} \!     
   \left( \frac{\mu^2}{-s_{ij}} \right)^{\!\epsilon} \!, \nonumber
\end{eqnarray}
\begin{eqnarray}
   \bm{I}^{(2)}(\epsilon)  
   &=& \frac{e^{-\epsilon\gamma_E}\,\Gamma(1-2\epsilon)}%
            {\Gamma(1-\epsilon)} 
    \left( K + \frac{\beta_0}{2\epsilon} \right) 
    \bm{I}^{(1)}(2\epsilon) \\
   &&\mbox{}- \frac12\,\bm{I}^{(1)}(\epsilon)  
    \left( \bm{I}^{(1)}(\epsilon) + \frac{\beta_0}{\epsilon} \right)  
    + \bm{H}_{\rm R.S.}^{(2)}(\epsilon) \,, \qquad \nonumber
\end{eqnarray}
where $s_{ij}\equiv 2\sigma_{ij}\,p_i\cdot p_j+i0$, and the sign factor $\sigma_{ij}=+1$ if the momenta $p_i$ and $p_j$ are both incoming or outgoing, and $\sigma_{ij}=-1$ otherwise. The quantities $K$ and $g_i$ in (\ref{I2}) are related to anomalous-dimension coefficients given below as $K=\gamma_1^{\rm cusp}/(2\gamma_0^{\rm cusp})$ and $g_i=-\gamma_0^i/2$, while $\beta_0=\frac{11}{3}\,C_A-\frac{4}{3}\,T_F n_f$ is the first coefficient of the QCD $\beta$-function. The sums in the expressions above are over all external partons. We use the color-space formalism of \cite{Catani:1996jh}, in which $n$-particle amplitudes are treated as $n$-dimensional vectors in color space. $\bm{T}_i$ is the color generator associated with the $i$-th parton and acts as a matrix on its color index. The product $\bm{T}_i\cdot\bm{T}_j\equiv T_i^a\,T_j^a$ is summed over $a$. Generators associated with different particles trivially commute, $\bm{T}_i\cdot\bm{T}_j=\bm{T}_j\cdot\bm{T}_i$ for $i\ne j$, while $\bm{T}_i^2=C_i$ is given in terms of the quadratic Casimir operator of the corresponding color representation, i.e., $C_q=C_{\bar q}=C_F$ and $C_g=C_A$. Owing to color conservation, the scattering amplitudes fulfill the relation $\sum_i\,\bm{T}_i\,|{\cal M}_n(\epsilon,\{p\})\rangle = 0$. 

The scheme-dependent quantity $\bm{H}_{\rm R.S.}^{(2)}$, which only contains $1/\epsilon$ divergences, was not specified in \cite{Catani:1998bh} except for the simplest case of the on-shell quark form factor. In subsequent two-loop calculations of the three-parton $q\bar q g$ amplitude \cite{Garland:2001tf} and of $2\to 2$ scattering amplitudes \cite{Anastasiou:2000kg,Bern:2002tk} it was observed that its color-diagonal part has a universal structure. However, the calculations of four-parton amplitudes also revealed the presence of a non-trivial color structure \cite{Bern:2002tk}. A conjecture for the form of this term for a general $n$-parton amplitude was made in \cite{Bern:2004cz}. 

An interesting alternative approach to the problem of IR singularities of on-shell amplitudes was developed in \cite{Sterman:2002qn}, where the authors exploited the factorization properties of scattering amplitudes \cite{Sen:1982bt,Kidonakis:1998bk} along with IR evolution equations familiar from the analysis of the Sudakov form factor \cite{Magnea:1990zb}. They recovered Catani's result (\ref{I2}) at two-loop order and related the coefficient of the unspecified $1/\epsilon$ pole term to a soft anomalous-dimension matrix, which was unknown at the time. They also explained how their method could be extended beyond two-loop order. The two-loop soft anomalous-dimension matrix was later calculated in \cite{MertAybat:2006wq}, where its color structure was found to be proportional to that obtained at one-loop order. 

In this Letter we propose an {\em all-order\/} generalization of Catani's result (\ref{I2}) valid for an arbitrary on-shell $n$-parton scattering amplitude. We find that in a minimal subtraction scheme the color structure of the IR pole terms is simpler than anticipated based on Catani's work \cite{Catani:1998bh}. In fact, to all-loop order the $1/\epsilon$ pole term contains only the structures $\bm{1}$ and $\bm{T}_i\cdot\bm{T}_j$. Our analysis is based on effective field theory and shares many similarities with that of \cite{Sterman:2002qn}. However, in our case the hard, jet, and soft functions are defined in terms of matrix elements of different types of fields in the effective theory and are in one-to-one correspondence with different physical scales. The corresponding definitions in \cite{Sterman:2002qn} are less intuitive. 

Our key observation is that the IR singularities of on-shell amplitudes in massless QCD are in one-to-one correspondence to the UV poles of operator matrix elements in soft-collinear effective theory (SCET) \cite{Bauer:2000yr,Beneke:2002ph}. They can be subtracted by means of a multiplicative renormalization factor $\bm{Z}$ (a matrix in color space), whose structure is constrained by the renormalization group (RG). SCET is the appropriate effective theory to analyze scattering processes at large momentum transfer, which involve jets (or individual hadrons) with small invariant masses. It separates hard contributions associated with the large momentum transfer from low-energy contributions associated with the small invariant masses of the initial- and final-state particles. For a general $n$-jet observable, the effective theory involves a set of collinear fields for each direction of large energy flow, which describe the QCD dynamics inside the individual jets. It also contains soft quark and gluon fields, which mediate low-energy interactions among the jets. Hard interactions are integrated out and absorbed into the Wilson coefficients of operators built from soft and collinear fields. A generic $n$-jet process is mediated by an effective Hamiltonian ${\cal H}_n=\sum_i\,{\cal C}_{n,i}(\mu)\,O_{n,i}^{\rm ren}(\mu)$, where the sum runs over a basis of SCET operators built from $n$ distinct types of collinear fields. The bare matrix elements of these operators are UV divergent and are renormalized in the $\overline{\rm MS}$ scheme. Their divergences are absorbed into a renormalization factor via $O_{n,i}^{\rm ren}(\mu)=\sum_j Z_{ij}(\mu,\epsilon)\,O_{n,j}^{\rm bare}(\epsilon)$. For physical quantities, the scale dependence of the Wilson coefficients ${\cal C}_{n,i}(\mu)$ cancels against that of the matrix elements of the renormalized operators. 

In a physical process with initial- and final-state hadrons, the soft and collinear scales are set by nonperturbative dynamics or experimental cuts. Let us now consider (slightly) off-shell $n$-parton amputated Green's functions $G_n(\{p\})$. In this case the jet-scale $\Lambda_J^2$ is set by the off-shellness $p_i^2$ of the fields, and the soft scale is $\Lambda_s\sim\Lambda_J^2/Q$, where $Q$ is a typical hard momentum transfer. The Green's functions are obtained by taking matrix elements of the above effective Hamiltonian, which can be written as
\begin{equation}\label{greensfunction}
   G_n(\{p\}) = \lim_{\epsilon\to 0}\,\sum_{i,j}\,
   {\cal C}_{n,i}(\mu)\,Z_{ij}(\mu,\epsilon)\,
   \langle O_{n,j}^{\rm bare}(\epsilon) \rangle \,,
\end{equation} 
where we suppress the dependence of the quantities on the right-hand side on the parton momenta. To obtain on-shell $n$-parton scattering amplitudes from these Green's functions one takes the limit $p_i^2\to 0$. This introduces IR divergences, which can be regulated by evaluating the effective-theory matrix elements in $d=4-2\epsilon$ dimensions. Doing so renders the matrix elements of the operators $O_j^{\rm bare}$ trivial: in the limit $p_i^2\to0$ both the soft and the jet scales tend to zero, and all loop diagrams in the effective theory become scaleless and vanish. The bare matrix elements are thus reduced to trivial Dirac and color structures. Since the IR divergences are independent of the spin structure, we will not make the Dirac structures explicit but simply absorb them into the Wilson coefficients. The on-shell Green's functions are then directly proportional to the Wilson coefficients of $n$-jet SCET operators in the $\overline{\rm MS}$ scheme. In the color-space basis notation of (\ref{I2}), the effective Hamiltonian reads ${\cal H}_n=\langle O_n^{\rm ren}|{\cal C}_n\rangle$, and we have
\begin{equation}\label{renorm}
   |{\cal C}_n(\{p\},\mu)\rangle 
   = \lim_{\epsilon\to 0}\,\bm{Z}^{-1}(\epsilon,\{p\},\mu)\,
   |G_n(\epsilon,\{p\})\rangle \,.
 \end{equation}
This notation is convenient but unconventional, in that our Wilson coefficients and operators are not separately color singlets and Lorentz scalars. The scattering amplitudes $|{\cal M}_n(\epsilon,\{p\})\rangle$  are obtained by contracting the amputated on-shell Green's functions with the spinors and polarization vectors associated with the external particles. Their singularities are thus governed by the same $\bm{Z}$ matrix.

The logarithm of the renormalization factor $\bm{Z}$ in (\ref{renorm}) is related via $\bm{\Gamma}=-d\ln\bm{Z}/d\ln\mu$ to the anomalous-dimension matrix $\bm{\Gamma}$ governing the RG evolution of the $n$-jet SCET operators $O_n^{\rm ren}$. The same quantity controls the evolution of the Wilson coefficients, and hence of the minimally subtracted on-shell scattering amplitudes, via the evolution equation
\begin{equation}\label{RGE}
   \frac{d}{d\ln\mu}\,|{\cal C}_n(\{p\},\mu)\rangle
   = \bm{\Gamma}\,|{\cal C}_n(\{p\},\mu)\rangle \,.
\end{equation}
We will now present a conjecture for the exact form of the anomalous-dimension matrix. In general, $\bm{\Gamma}=\bm{\Gamma}_{c+s}$ is determined by the anomalous-dimension contributions of collinear and soft modes in the SCET matrix elements. An important feature of SCET is that the interactions of collinear fields with soft gluons can be removed by field redefinitions and absorbed into soft Wilson lines \cite{Bauer:2000yr}. Interactions with soft quarks are power suppressed and can be ignored. Moreover, the different collinear sectors in SCET do not interact with each other. This allow us to decompose $\bm{\Gamma}=\bm{\Gamma}_s+\sum_i\gamma_c^i$, where the one-particle collinear contributions are diagonal in color space. Hence, contributions to the anomalous dimension involving correlations between several partons only reside in the soft contribution $\bm{\Gamma}_s$. After the decoupling transformation the soft matrix element is a vacuum expectation value $\langle 0|\bm{S}_1\ldots\bm{S}_n|0\rangle$ of $n$ light-like Wilson lines, one for each external parton (see also \cite{Kidonakis:1998bk}). Here
\begin{equation}\label{softwilson}
   \bm{S}_i = {\bf P} \exp\left[ ig \int_{-\infty}^0\!dt\,
   n_i\cdot A^a(t n_i)\,T_i^a \right] ,
\end{equation}
where $n_i$ is a light-like reference vector aligned with the $i$-th particle's momentum. When the color indices are contracted into color-singlet combinations, the soft matrix element consists of one or more Wilson loops (closed at infinity), whose lines cross and or have cusps at the origin. The renormalization properties of Wilson loops have been studied extensively in the literature (see e.g.\ \cite{Brandt:1981kf,Frenkel:1984pz,Korchemsky:1985xj}). Based on these results and on other considerations to be explained in detail elsewhere, we propose that the exact expression for the anomalous-dimension matrix of an $n$-jet operator in SCET in the color-space formalism is 
\begin{equation}\label{magic}
   \bm{\Gamma} 
   = \sum_{(i,j)}\,\frac{\bm{T}_i\cdot\bm{T}_j}{2}\,\,
   \gamma_{\rm cusp}(\alpha_s)\,\ln\frac{\mu^2}{-s_{ij}} 
    + \sum_i\,\gamma^i(\alpha_s) \,.
\end{equation}
The sums run over the $n$ external partons. Here and below the notation $(i_1,...,i_k)$ refers to unordered tuples of distinct parton indices. Our result contains three universal anomalous-dimension functions. The quantity $\gamma_{\rm cusp}$ is proportional to the cusp anomalous dimension of Wilson loops with light-like segments \cite{Korchemsky:1985xj}, while $\gamma^q\equiv\gamma^{\bar q}$ and $\gamma^g$ are anomalous dimensions specific for (anti-)quark and gluon fields. They are defined by relation (\ref{magic}). 

Our conjecture for the anomalous-dimension matrix is the simplest expression consistent with existing calculations. It implies that the cusp anomalous dimension characterizes the renormalization of Wilson lines even in the general case, where several lines meet at a single space-time point. The structure of the logarithmic terms in (\ref{magic}) agrees with an explicit two-loop calculation in \cite{MertAybat:2006wq}. For the quark and gluon form factors, the divergent terms are known to three-loop accuracy \cite{Moch:2005id}. When applied to this case, our general relation (\ref{magic}) implies that the cusp anomalous dimensions in the fundamental and adjoint representations of $SU(N_c)$ are related by $\Gamma_{\rm cusp}^F(\alpha_s)/C_F=\Gamma_{\rm cusp}^A(\alpha_s)/C_A\equiv\gamma_{\rm cusp}(\alpha_s)$. This relation is indeed fulfilled to three-loop order \cite{Moch:2004pa}. The application to the form factors also determines the anomalous dimensions $\gamma^q\equiv\gamma^{\bar q}$ and $\gamma^g$ to three-loop accuracy. In the effective theory the form factors are mapped onto two-jet operators containing two collinear quark or gluon fields. The corresponding three-loop anomalous dimensions were given in \cite{Becher:2007ty}. In the notation of these papers, we have $2\gamma^q(\alpha_s)=\gamma^V(\alpha_s)$ and $2\gamma^g(\alpha_s)=\gamma^t(\alpha_s)+\gamma^S(\alpha_s)+\beta(\alpha_s)/\alpha_s$, where $\beta(\alpha_s)=d\alpha_s/d\ln\mu$. These results are valid in the standard dimensional regularization scheme with $d$-dimensional helicities.

The relation $\bm{\Gamma}=-d\ln\bm{Z}/d\ln\mu$ may be integrated to obtain a closed expression for the logarithm of $\bm{Z}$. Using the relation $\beta(\alpha_s,\epsilon)=\beta(\alpha_s)-2\epsilon\,\alpha_s$ for the $\beta$-function in $d=4-2\epsilon$ dimensions, we find 
\begin{equation}\label{lnZ}
  \ln\bm{Z} 
   = - \int\limits_0^{\alpha_s} 
   \frac{d\alpha}{\beta(\alpha,\epsilon)}\,\Bigg[ \bm{\Gamma}(\alpha) 
   + \Gamma'(\alpha) \int\limits_{\alpha_s}^\alpha 
   \frac{d\alpha'}{\beta(\alpha',\epsilon)} \Bigg] \,,
\end{equation}
where we have defined
\begin{equation}
   \Gamma' = \frac{\partial}{\partial\ln\mu}\,\bm{\Gamma}
   = - \gamma_{\rm cusp}(\alpha_s)\,\sum_i\,C_i \,.
\end{equation}
When integrating over $\alpha$ in (\ref{lnZ}), the scale $\mu$ in the argument of the logarithm in $\bm{\Gamma}$ must be kept fixed. Note that when acting on color-singlet states, the unweighted sum over color generators satisfies the relation 
\begin{equation}\label{colorrel}
   \sum_{(i,j)}\,\bm{T}_i\cdot\bm{T}_j
   = - \sum_i\,\bm{T}_i^2 = - \sum_i\,C_i \,,
\end{equation}
which follows from color conservation. Since both the scattering amplitudes and the anomalous dimension $\bm{\Gamma}$ are color-conserving, this relation can always be used in our case. It is understood that the result (\ref{lnZ}) must be expanded in powers of $\alpha_s$ with $\epsilon$ treated as a fixed ${\cal O}(\alpha_s^0)$ quantity. Writing the perturbative series of the anomalous dimension and $\beta$-function in the form $\bm{\Gamma}(\alpha_s)=\sum_n\bm{\Gamma}_n\,(\frac{\alpha_s}{4\pi})^{n+1}$ and $\beta(\alpha_s)=-2\alpha_s\sum_n\beta_n\,(\frac{\alpha_s}{4\pi})^{n+1}$, we find up to three-loop order the result
\begin{eqnarray}\label{result}
   \ln\bm{Z} 
   &=& \frac{\alpha_s}{4\pi} 
    \left( \frac{\Gamma_0'}{4\epsilon^2}
    + \frac{\bm{\Gamma}_0}{2\epsilon} \right) \nonumber\\
   &&\hspace{-6.3mm}\mbox{}+ \left( \frac{\alpha_s}{4\pi} \right)^2 \! 
    \left[ - \frac{3\beta_0\Gamma_0'}{16\epsilon^3} 
    + \frac{\Gamma_1'-4\beta_0\bm{\Gamma}_0}{16\epsilon^2}
    + \frac{\bm{\Gamma}_1}{4\epsilon} \right] \nonumber\\
   &&\hspace{-6.3mm}\mbox{}+ \left( \frac{\alpha_s}{4\pi} \right)^3 
    \Bigg[ \frac{11\beta_0^2\,\Gamma_0'}{72\epsilon^4}
    - \frac{5\beta_0\Gamma_1' + 8\beta_1\Gamma_0' 
            - 12\beta_0^2\,\bm{\Gamma}_0}{72\epsilon^3} \nonumber\\
   &&\quad\mbox{}+ 
    \frac{\Gamma_2' - 6\beta_0\bm{\Gamma}_1 
          - 6\beta_1\bm{\Gamma}_0}{36\epsilon^2}
    + \frac{\bm{\Gamma}_2}{6\epsilon} \Bigg] 
    + \dots \,,
\end{eqnarray}
which only contains the two color structures present in (\ref{magic}). Exponentiation yields an explicit expression for $\bm{Z}$, which simplifies since the different expansion coefficients $\bm{\Gamma}_n$ commute. The relevant one-loop coefficients of the three anomalous-dimension functions are $\gamma_0^{\rm cusp}=4$, $\gamma_0^q=-3C_F$, and $\gamma_0^g=-\beta_0$. The highest pole in the ${\cal O}(\alpha_s^n)$ term of $\ln\bm{Z}$ is $1/\epsilon^{n+1}$, instead of $1/\epsilon^{2n}$ for the $\bm{Z}$ factor itself. The exponentiation of the higher pole terms was noted previously in \cite{Sterman:2002qn}.

The one- and two-loop coefficients of the matrix $\bm{Z}$ are closely related to Catani's subtraction operators $\bm{I}^{(1)}$ and $\bm{I}^{(2)}$ given in (\ref{I2}). The conditions linking his objects to ours are that the differences $2\bm{I}^{(1)}-\bm{Z}_1$ and $4\bm{I}^{(2)}+2\bm{I}^{(1)}\bm{Z}_1-\bm{Z}_2$ remain finite for $\epsilon\to 0$. Here $\bm{Z}_n$ denotes the coefficient of $(\alpha_s/4\pi)^n$ in the $\bm{Z}$ factor. The first relation is indeed satisfied. The second one can be used to derive an explicit expression for the quantity $\bm{H}_{\rm R.S.}^{(2)}$ in~(\ref{I2}) entering the two-loop coefficient of the $1/\epsilon$ pole, which was not obtained in \cite{Catani:1998bh}. We find
\begin{eqnarray}\label{H2}
   \bm{H}_{\rm R.S.}^{(2)} 
   =&& \frac{1}{16\epsilon}\,\sum_i \left( \gamma_1^i 
    - \frac{\gamma_1^{\rm cusp}}{\gamma_0^{\rm cusp}}\,\gamma_0^i
    + \frac{\pi^2}{16}\,\beta_0\,\gamma_0^{\rm cusp}\,C_i \right) \\
   &&\hspace{-0.5cm}+ \frac{if_{abc}}{24\epsilon}\,
    \sum_{(i,j,k)} T_i^a\,T_j^b\,T_k^c\,
    \ln\frac{-s_{ij}}{-s_{jk}} \ln\frac{-s_{jk}}{-s_{ki}} 
    \ln\frac{-s_{ki}}{-s_{ij}} \nonumber \\
  &&\hspace{-1.0cm}
     - \frac{if^{abc}}{32\epsilon}\sum_{(i,\,j,\,k)} T_i^a\, T_j^b\,T_k^c\;
\bigg( \frac{\gamma_0^i}{C_i} - \frac{\gamma_0^j}{C_j} \bigg)
\ln\frac{-s_{ij}}{-s_{jk}}\,\ln\frac{-s_{ki}}{-s_{ij}}  \,, \nonumber
\end{eqnarray}
which apart from the last  two terms is diagonal in color space and universal in the sense that it is a sum over contributions from individual partons. Note that only the first term in this result is of a form suggested by (\ref{result}). The remaining terms in the first line arise because two-loop corrections involving the cusp anomalous dimension or the $\beta$-function are not implemented in an optimal way in (\ref{I2}). More importantly, the last two terms in (\ref{H2}), which are present for four or more partons, arise only because the subtraction operators $\bm{I}^{(k)}$ in \cite{Catani:1998bh} are not defined in a minimal scheme but also include ${\cal O}(\epsilon^k)$ terms with $k\ge 0$. 

Our expressions (\ref{result}) and (\ref{H2}) reproduce all known results for the two-loop $1/\epsilon^k$ poles of on-shell scattering amplitudes in massless QCD. In addition to the on-shell quark and gluon form factors, these include $e^+e^- \to \bar qqg$ \cite{Garland:2001tf} as well as all four-point functions of quarks and gluons \cite{Anastasiou:2000kg,Bern:2002tk}. It further confirms the ansatz made for higher-point functions in \cite{Bern:2004cz}. At the three-loop level, only the IR divergences of the quark and gluon form factors are known for the QCD case \cite{Moch:2005id}. For ${\cal N}=4$ supersymmetric Yang-Mills theory in the planar limit, on the other hand, the four-point functions are known up to four-loop order \cite{Anastasiou:2003kj}. The divergent part of these amplitudes factors into a product of square roots of form factors of neighbouring legs, which is consistent with the structure of our anomalous dimension, given that at leading order in $1/N_c$ one has $\bm{T}_i\cdot\bm{T}_j\to-\frac{N_c}{2}\,\bm{1}$ for neighbouring legs and zero otherwise. A more stringent test of our conjecture could be performed using existing three-loop results for the full four-gluon scattering amplitude in this theory \cite{Bern:2008pv}, if one succeeds in evaluating the non-planar master integrals.

The most important application of our result is the resummation of Sudakov logarithms in $n$-jet processes. The evolution equation (\ref{RGE}) is simple enough to admit exact solutions in closed form, and the known three-loop anomalous dimensions allow for resummations at next-to-next-to-next-to-leading logarithmic accuracy. Our formalism can be generalized to processes involving massive partons by combining it with methods developed in \cite{Mitov:2006xs,Becher:2007cu}. The great simplicity of our result (\ref{magic}) appears to hint at a universal origin of IR singularities that is disconnected from the genuine dynamics of the scattering amplitude itself.

{\em Acknowledgments:\/}
We are grateful to N.~Arkani-Hamed, L.~Dixon and J.~Maldacena for useful comments. T.B.\ was supported by the U.S.\ DOE under Grant DE-AC02-76CH03000. Fermilab is operated by the Fermi Research Alliance under contract with the DOE.


\begin{thebibliography}{99}

\bibitem{Kinoshita:1962ur}
  T.~Kinoshita,
  J.\ Math.\ Phys.\  {\bf 3}, 650 (1962);
  T.~D.~Lee and M.~Nauenberg,
  Phys.\ Rev.\  {\bf 133}, B1549 (1964).
  
\bibitem{Yennie:1961ad}
  D.~R.~Yennie, S.~C.~Frautschi and H.~Suura,
  Annals Phys.\  {\bf 13}, 379 (1961).
  
 \bibitem{Frenkel:1984pz}
  J.~G.~M.~Gatheral,
  Phys.\ Lett.\  B {\bf 133}, 90 (1983);
  J.~Frenkel and J.~C.~Taylor,
  Nucl.\ Phys.\  B {\bf 246}, 231 (1984).

\bibitem{Collins:1989gx}
  J.~C.~Collins, D.~E.~Soper and G.~Sterman,
  Adv.\ Ser.\ Direct.\ High Energy Phys.\  {\bf 5}, 1 (1988).
    
\bibitem{Catani:1998bh}
  S.~Catani,
  Phys.\ Lett.\  B {\bf 427}, 161 (1998).

\bibitem{Catani:1996jh}
  S.~Catani and M.~H.~Seymour,
  Phys.\ Lett.\  B {\bf 378}, 287 (1996); 
%
  Nucl.\ Phys.\  B {\bf 485}, 291 (1997)
  [Erratum-ibid.\  B {\bf 510}, 503 (1998)].

\bibitem{Garland:2001tf}
  L.~W.~Garland et al., 
  Nucl.\ Phys.\  B {\bf 627}, 107 (2002).
    
\bibitem{Anastasiou:2000kg}
  C.~Anastasiou, E.~W.~N.~Glover, C.~Oleari and M.~E.~Tejeda-Yeomans,
  Nucl.\ Phys.\  B {\bf 601}, 318 (2001); 
%
  Nucl.\ Phys.\  B {\bf 601}, 341 (2001);
%
  Nucl.\ Phys.\  B {\bf 605}, 486 (2001);
%
  E.~W.~N.~Glover, C.~Oleari and M.~E.~Tejeda-Yeomans,
  Nucl.\ Phys.\  B {\bf 605}, 467 (2001).

\bibitem{Bern:2002tk}
  Z.~Bern, A.~De Freitas and L.~J.~Dixon,
  JHEP {\bf 0203}, 018 (2002); 
%
  JHEP {\bf 0306}, 028 (2003).

\bibitem{Bern:2004cz}
  Z.~Bern, L.~J.~Dixon and D.~A.~Kosower,
  JHEP {\bf 0408}, 012 (2004).
  
\bibitem{Sterman:2002qn}
  G.~Sterman and M.~E.~Tejeda-Yeomans,
  Phys.\ Lett.\  B {\bf 552}, 48 (2003).
  
\bibitem{Sen:1982bt}
  A.~Sen,
  Phys.\ Rev.\  D {\bf 28}, 860 (1983).
  
\bibitem{Kidonakis:1998bk}
  N.~Kidonakis, G.~Oderda and G.~Sterman,
  Nucl.\ Phys.\  B {\bf 525}, 299 (1998);
%
  Nucl.\ Phys.\  B {\bf 531}, 365 (1998).
  
\bibitem{Magnea:1990zb}
  L.~Magnea and G.~Sterman,
  Phys.\ Rev.\  D {\bf 42}, 4222 (1990).

\bibitem{MertAybat:2006wq}
  S.~Mert Aybat, L.~J.~Dixon and G.~Sterman,
  Phys.\ Rev.\ Lett.\  {\bf 97}, 072001 (2006);
%
  Phys.\ Rev.\  D {\bf 74}, 074004 (2006).
  
\bibitem{Bauer:2000yr}
  C.~W.~Bauer, S.~Fleming, D.~Pirjol and I.~W.~Stewart,
  Phys.\ Rev.\  D {\bf 63}, 114020 (2001);
%
  C.~W.~Bauer, D.~Pirjol and I.~W.~Stewart,
  Phys.\ Rev.\  D {\bf 65}, 054022 (2002).

\bibitem{Beneke:2002ph}
  M.~Beneke, A.~P.~Chapovsky, M.~Diehl and T.~Feldmann,
  Nucl.\ Phys.\  B {\bf 643}, 431 (2002).

\bibitem{Brandt:1981kf}
  R.~A.~Brandt, F.~Neri and M.~A.~Sato,
  Phys.\ Rev.\  D {\bf 24}, 879 (1981);
%
  R.~A.~Brandt, A.~Gocksch, M.~A.~Sato and F.~Neri,
  Phys.\ Rev.\  D {\bf 26}, 3611 (1982).

\bibitem{Korchemsky:1985xj}
  G.~P.~Korchemsky and A.~V.~Radyushkin,
  Phys.\ Lett.\  B {\bf 171}, 459 (1986);
%
  Nucl.\ Phys.\ B {\bf 283}, 342 (1987);
%
  I.~A.~Korchemskaya and G.~P.~Korchemsky,
  Phys.\ Lett.\ B {\bf 287}, 169 (1992).

\bibitem{Moch:2005id}
  S.~Moch, J.~A.~M.~Vermaseren and A.~Vogt,
  JHEP {\bf 0508}, 049 (2005);
%
  Phys.\ Lett.\  B {\bf 625}, 245 (2005).
  
\bibitem{Moch:2004pa}
  S.~Moch, J.~A.~M.~Vermaseren and A.~Vogt,
  Nucl.\ Phys.\  B {\bf 688}, 101 (2004).

\bibitem{Becher:2007ty}
  T.~Becher, M.~Neubert and G.~Xu,
  JHEP {\bf 0807}, 030 (2008);
%
  V.~Ahrens, T.~Becher, M.~Neubert and L.~L.~Yang,
  Eur.\ Phys.\ J.\  C {\bf 62}, 333 (2009).

\bibitem{Anastasiou:2003kj}
  C.~Anastasiou, Z.~Bern, L.~J.~Dixon and D.~A.~Kosower,
  Phys.\ Rev.\ Lett.\  {\bf 91}, 251602 (2003);
%
  Z.~Bern, L.~J.~Dixon and V.~A.~Smirnov,
  Phys.\ Rev.\  D {\bf 72}, 085001 (2005);
%
  Z.~Bern et al., 
  Phys.\ Rev.\  D {\bf 75}, 085010 (2007).

\bibitem{Bern:2008pv}
  Z.~Bern et al., 
  Phys.\ Rev.\  D {\bf 78}, 105019 (2008).

\bibitem{Mitov:2006xs}
  A.~Mitov and S.~Moch,
  JHEP {\bf 0705}, 001 (2007).

\bibitem{Becher:2007cu}
  T.~Becher and K.~Melnikov,
  JHEP {\bf 0706}, 084 (2007).

\end{thebibliography}
\end{document}